# UNCERTAINTY AND THE COMMUNICATION OF TIME

Loet Leydesdorff

Department of Science and Technology Dynamics

Nieuwe Achtergracht 166

1018 WV   AMSTERDAM

The Netherlands

**Abstract**

Prigogine and Stengers (1988) [47] have pointed to the centrality of the concepts of "time and eternity" for the cosmology contained in Newtonian physics, but they have not addressed this issue beyond the domain of physics.   The construction of "time" in the cosmology dates back to debates among Huygens, Newton, and Leibniz.   The deconstruction of this cosmology in terms of the philosophical questions of the 17th century suggests an uncertainty in the time dimension.   While order has been conceived as an "harmonie préétablie," it is considered as emergent from an evolutionary perspective.   In a "chaology", one should fully appreciate that different systems may use different clocks.   Communication systems can be considered as contingent in space *and* time: substances contain force or action, and they communicate not only in (observable) extension, but also over time.   While each communication system can be considered as a system of reference for a special theory of communication, the addition of an evolutionary perspective to the mathematical theory of communication opens up the possibility of a general theory of communication.

# UNCERTAINTY AND THE COMMUNICATION OF TIME

**Introduction**

In 1690, Christiaan Huygens noted that:
> "(I)t is not well to identify certitude with clear and distinct perception, for it is evident that there are, so to speak, various degrees of that clearness and distinctness. We are often deluded in things which we think we certainly understand. Descartes is an example of this; it is so with his laws of communication of motion by collision of bodies."[i]

Huygens made this distinction between clarity and certitude primarily because he wished to emphasize the importance of experimental work.[ii] However, this methodological critique of Descartes' ideas has an epistemological implication. If clarity and certainty are not necessarily related, they are different dimensions of knowledge: clarity in knowledge should be opposed to confusion, and certainty to uncertainty. The dynamics between these two dimensions of knowledge merit further specification. The problem of a dynamic interaction, however, presumes a notion of time. Indeed, the construction of "time" has been crucial to the development of the new philosophy in the second half of the 17th century.

The problem of the communication of time among systems (e.g., clocks) was central to Huygens' research programme; the differential calculus enabled Newton and Leibniz to develop the concepts of infinite and continuous time within the new physics. Towards the end of the 17th century, these scholars provided natural philosophy with firm mathematical and metaphysical foundations. Additionally, Newton and Huygens formulated methodologies on how to achieve more clarity and certainty by empirical investigations.

On the one hand, Newton tended towards the empiricist position when he formulated his well-known "*hypotheses non fingo*":
> "But hitherto I have not yet been able to discover the cause of those properties of gravity from phenomena, and I frame no hypotheses; for whatever is not deduced from the phenomena is to be called an hypothesis; and hypotheses, whether metaphysical or physical, whether of occult qualities or mechanical, have no place in experimental philosophy. In this philosophy particular propositions are inferred from the phenomena, and afterwards rendered general by induction." ([44], at p. 547.)

On the other hand, Huygens provided us with the rationalist counter-position in the *Cosmotheoros* (1698):
"I must acknowledge that what I here intend to treat of is not of that Nature as to admit of a certain knowledge; I cannot pretend anything as posititvely true (for how is it possible), but only to advance a probable Guess, the Truth of which everyone is at his own liberty to examine."[iii]

These two positions have more or less structured the discussion about scientific methods over the past centuries. The two positions, however, have in common a firm belief that one can take either the (un)certainty on the side of the objects of study or the (un)clarity in the analyst's mind, and from that starting point unambiguouly bridge the gap between the world and our understanding of the world, since a pre-established correspondence between the two can be assumed as the basis for their interaction.[32]

Further reflections in philosophy on the nature of this transcendental assumption have affected the development of physics only marginally, since for physics the epistemological boundaries of the Newtonian cosmology remained largely unproblematic.[iv] For example, Einstein and Infeld acknowledged this cosmology in 1938 as follows:
"Without the belief that it is possible to grasp the reality with our theoretical constructions, without the belief in the inner harmony of our world, there could be no science." ([13], at p. 296.)

Other natural scientists (e.g., [8, 46]) have discussed "the arrow of time," but they retained the idea of "a unified vision of time."[8]

Philosophical reflections, however, have been important for the social sciences, since there are many possible understandings of the social world, and many social worlds. In this context, "reality" and its harmony can no longer be taken for granted. As soon as there are more than two systems to synchronize, the interaction can in principle be decomposed in more than one way, and therefore the transcendental relation may itself become uncertain. If this is historically reflected in philosophy--as it has been--the issue is no longer whether one should build upon the bank of subjective (un)clarity or on the (un)certainty in the phenomena, but rather the question of which uncertainty or which unclarity one may wish and/or be able to build on.   In the absence of a single metaphysical guarantee for preestablished harmony and cosmos, asynchronicity will prevail.

In this study, I first deconstruct the modern cosmology in terms of the philosophical questions which have been basic to the mathematization of physics in the 17th century.   The deconstruction of the cosmology suggests an uncertainty in the time dimension.   In the second part of the study, I shall argue that one can nowadays specify the conditions under which clarity can be generated in relations among systems which contain and process uncertainties.   Since the concept of uncertainty can now be mathematically defined,[51] various problems of the 17th century can be reformulated.   For example, uncertainty can be considered as the substance of communication.   Communication systems can be studied in space *and* time: they operate in terms of substances which should be considered as force or action.   Observed harmony between substances requires explanation.

1. **The construction of the modern cosmology**
1.1. **"Uncertainty" in the New Philosophy**

According to Descartes, the act of doubt provides us with a point of departure for further investigations.   One is able to infer reflexively from the uncertainty which one finds in one's Ego ("*cogito*") to clarity concerning the existence of the subject of this reflection ("*ergo sum*").   With hindsight, Huygens' analysis clarified that Descartes had formulated a one-dimensional theory of knowledge, namely one in which the subject is able to replace uncertainty with clarity by reflection.   In order to be able to distinguish between mathematical clarity and empirical uncertainty, Huygens needed a two-dimensional theory of knowledge: whatever one derives on *a priori* grounds, and however clear this may be in mathematical terms, the inference remains an hypothesis about the physical world which yet needs to be tested empirically in order to become more certain.

What is the nature of the relation between contingent uncertainty and analytical clarity if one distinguishes between the two? Let me quote Huygens again:

> "Against Cartesius' dogma, that the nature or notion of a body should consist in extension alone, I have a notion of space that differs from the notion of a body: space is what may be occupied by a body."[v]

Note that Huygens uses "may be." In opposition to Descartes, this natural philosopher and his contemporaries had achieved an understanding of *empty* space (Newton) and *infinite* time (Leibniz), which allowed them to use abstract mathematical theories to draw inferences about physical reality which could be tested. Thus, the arrow of inference was reversed: space was no longer considered exclusively as a consequence of the extension of matter, as had been the case in Cartesian philosophy. Newton would radicalize this point of view and introduce concepts like gravity, which cannot easily be given a geometrical interpretation, while the availability of such (algebraic) concepts is even conditional for the physical understanding.

While Newtonian thought is most versatile in terms of an idealized mathematical system in addition to the contingent mechanical worldview, the Cartesian Huygens was pursued by philosophical problems. Huygens, however, was in the first place a physicist; he was so deeply impressed by Newton's *Principia* (1687) that he expressed the wish to pay the author a visit, which became possible after the Glorious Revolution in England (1688-1689). After his return he stated in a letter to his friend Leibniz that he found Newton's hypothesis concerning gravitation still "absurd."[vi] Analogously, he had reservations concerning Leibniz' differential notations, since they were based on algebra and not on geometry. However, from 1690 onwards, Huygens began to use Leibniz' notation for differentials along with ideas from Newton's physics in his own work, despite his philosophical reservations. Physics had definitively become one theoretical system.

## 1.2. **The Assumption of A System of Reference**

The philosophical point in the above quotations is different from the question of their usefulness for the understanding of the history of early modern physics. Obviously, the "*cogito*" leaves room for other notions of the "*res extensa*" than the Cartesian identification of a body with extension. If one is uncertain, one is uncertain *about* something. But is the *cogito* itself able to determine also the nature of the *res extensa*?

The *cogito* itself clarifies only the contingency of the *cogitans*: a system which is in doubt about itself is reflexively aware that it could have been otherwise, i.e., that it is contingent. This contingency refers to other possible states of the same system. The system which is uncertain, refers to a demarcation from something else (e.g., itself in another state) which can thus be considered as environment. But a reference to a demarcation is not a demarcation! In the act of doubt, the contingency cannot determine itself substantively, since it does not in itself contain knowledge about the existence or the nature of an outside world.

Therefore, the theory of knowledge in Cartesian philosophy remained internal to the *Ego*. The argument of "*Cogito ergo sum*" preceded the step in which Descartes invoked the Goodness of God ("*Veracitas Dei*") as a warrant that our (internal) imaginings about the (external) world correspond with a physical reality (including our own corporal existence). There is nothing in contingency itself which guarantees that this environment exists as "*res extensa*," i.e., as physical matter, and not as mere imagination. The self-reference, however, provides the reflexive *cogito* with a previous state, and thus with a reference to *finite* time. Consequently, the delineation of the *contingent Ego* implies a reference to a *transcendent Other*, which is expected to contain infinite time. However, the contingent self can only be delineated negatively from its Transcendency. The Transcendency remains only an expectation. Any positive delineation of the contingency requires additional information, i.e., information which does not originate internally within the *cogito*, but from its relation with an environment.

As long as there is no delineation from an external system, there can only be contingency in relation to transcendency.[vii] As soon as something else is considered as different but contingent, one has to assume communication between system and environment, communication in time, and communication of the system's time.

### 1.3. "Time" in the New Philosophy

The question about how time is communicated among systems and with reference to infinite time, therefore, was crucial to the new philosophy. In relation to transcendency, contingency contained only its own time which could negatively be delineated from infinite time, i.e. Eternity. In order to infer beyond God to the existence of a contingent system other than the *cogito*, one had to raise the question of how the systems manage to remain synchronous over time. Can they use their mutual communications for updates or do they have to refer independently to a "standard clock"? Is it necessary to specify God's role in the synchronization among the substances?

In philosophy the synchronicity problem is at the core of the well-known mind-body problem: how do the body and the mind communicate when knowledge of the physical world is generated, and subsequently, how do they communicate in human action as an expression of the free will? Descartes originally raised this question in terms of the communication between the substances: how do the *res cogitans* (thinking) and the *res extensa* (matter) communicate? The metaphor of two clocks which run synchronously was introduced by the Cartesian Geulincx. However, not only the metaphor, but also the formulation in terms of *communication* between two systems remained central throughout the 17th century. For example, when Leibniz published his system in the *Journal des Savants* at the end of this century, he entitled his treatise "New systems of the nature and of the communication of substances, and of the union between the soul and the body" [32].

In the metaphor of the synchronicity between two clocks, the one clock represents the physical world, the other the spiritual one. How does it happen that our mental perceptions correspond with reality? As noted, Descartes' metaphysical answer to this problem had been that the Goodness of God implies that He is not expected to continuously deceive us. However, in a mechanistic philosophy, one would like an answer to the question of how this mechanism works also in physical terms.

Huygens made this very question central to his research programme for the new physics. The practical question of the day was the problem of keeping clocks synchronous on board of ships at sea. Huygens generalized this problem to the question of the communication between oscillating bodies in a study of 1673, entitled *Horologium oscillatorum*. Note that this latter study was not a contribution to the practical problem, which had already been amply discussed in his 1655 study entitled *Horologium*, but more importantly to the major theoretical problems in the new Natural Philosophy.[60]

While Huygens gave an essentially mechanistic answer to the question of how different systems communicate time, Geulincx at Leuven had proposed that at the moment of each communication God had to intervene to keep the two clocks synchronous (so-called "occasionalism"). In a study, entitled *Harmonie préétablie* (1696), Leibniz elaborated a third possibility for keeping the two clocks operating synchronically:

> "One may think of two clocks which are completely synchronous. This can only happen in three ways: firstly, it may be based on a mutual influence among them; secondly, that continuously somebody takes care; thirdly, on the internal precision of each of them." ([33], at p. 272.)

Leibniz then attributed the first hypothesis to Huygens; the second refers to the noted continuous need for intervention; and he chose the third option himself. This option enabled Leibniz to integrate into a single encompassing system the metaphysical issues at stake, the mathematical concept of infinite time which he (and Newton) had derived a few decades earlier when developing the calculus, and the mechanistic world picture of Cartesianism.

## 1.4. "The Time of the Lord is the Best of All Times"[viii]

Leibniz, however, emphasized the *hypothetical* character of the *harmonie préétablie*, which he proposed.   He formulated that

> "Once one has understood the possibility of this *hypothesis of correspondence*, one also understands that it accounts best for reason, and that it provides us with a wonderful image of the harmony in the universe, and of the perfection of God's works."[ix]

This meta-physical hypothesis addressed, among other things, a problem which had remained an open question within the mechanistic philosophy, namely how the human soul once embodied could return to the transcendent Eternity from which it was derived as a contingency in the inference as discussed above.   In addition to other radical implications (e.g., Spinozism), Cartesianism implied a mechanistic cosmology that could lead to contradictions in the basic assumptions concerning this issue in Christianity [24].   For example, we know from correspondence that Huygens was sometimes deeply troubled about the problem of the immortality of the soul.[x]

The quest for an encompassing solution became particularly urgent in 1685 when Protestantism was under vehement attack by the counter-reformation.   In this year, Louis XIV reinvoked the Edict of Nantes, and in England, a Catholic king (James II) had acceded to the throne.   Protestantism was on the defensive; one might even say on the verge of a breakdown.   Could it be provided with other options than a retreat to defensive orthodoxy in its relation to the new philosophy?   How could the internal contradictions between the new religion and the new philosophy be resolved in order to maintain both freedom of religion and the explaining power of the emerging modern science?   Was there any possibility of bringing these great systems into harmony?

In the winter of 1685-1686, Leibniz wrote the first draft of his *Discours de la Métaphysique*; Newton completed his *Principia*,[xi] to be published in 1687; and Huygens was ill and depressed in The Hague, since he was not allowed to return to the Academy in Paris of which he had been director for so many years.[xii]   Although there would remain differences of opinion among these three scholars,[xiii] in the years 1685-1689 the integrated system in terms of Newtonian physics, the calculus, and Protestant metaphysics was put into place. When Huygens came to visit Newton in 1689, his oldest brother Constantijn was the private secretary of the new King of England (William of Orange).   Newtonianism could thus become the basic ideology for the English revolution from 1689 onwards [23].   A metaphysically, scientifically, and ideologically supported coalition could be formed between England, Holland, and Prussia, which laid the foundation for the 18th-century Enlightenment.[xiv]

In the decades preceding these events, the various ingredients to resolve the tensions between the mechanistic philosophy and the Christian religion had been developed piecemeal in the relations and oppositions among Huygens, Leibniz, and Newton (see, e.g., [10]). Huygens agreed with Newton about replacing the Cartesian vortices with a concept of continuous and empty space; Leibniz and Newton had developed the mathematical idealization of differential calculus independently of each other; and all three of them believed in the existence of absolute and infinite time.   The grand synthesis, however, was forced by the historical situation.

After 1689, the scientific system had been brought into harmony with its surrounding culture by assuming order in the time dimension.   The human soul has to live on earth, i.e., in finite time, but its immortality provides it with the possibility to follow Christ, and to return to God's eternal time.[xv]   The semantics of differential calculus serves most graciously and convincingly to illustrate the transition between the transcendent and the contingent: the discreteness of this contingent world should be understood as a manifestation of continuous time and space.   The infinitesimal transition exhibits how worlds other than the one which we perceive with our senses resound within it.   One would not even be able to understand the contingent properly without drawing upon the idealized model.   More generally, understanding physical communication through the mathematical model provided a mental model to reconcile the idealistic and the mechanistic interpretations of experimental facts.

2. **The deconstruction of the modern cosmology**

The cosmology warranted order within each of the substances and between them, so that what seemed at first to be different (i.e., the Word and the world) could be resolved into harmonic correspondence.  The harmonic solution at the cosmological level warrants reconciliation at the metaphysical one: nature is revealed to us by God's grace, and therefore we are able to reconcile our mathematical image with physical reality.  While there is initially a gap between the complexity of the contingencies and the idealization in the model system, the two dimensions of mathematical clarity and empirical uncertainty can be brought to interact, and we are warranted in achieving scientific understanding, i.e., true knowledge about the world.

Thus, I showed that the question of how clarity can be related to uncertainty was raised in the 17th century, but was then answered in a specific way in order to secure the progress of physics in a non-secularized world.  I shall argue in the second part of this study that one can nowadays specify the conditions under which clarity can be generated in relations among systems which contain and process uncertainties.

2.1. **"Uncertainty" as the substance of communication**

Indeed, in the philosophy of science, in the social sciences, and most pronouncedly in the reflexive sociology of science (e.g., [59]), we have increasingly lost all notion of truth in the transcendental sense of fundamental certainty; we have become fundamentally uncertain.  Can anything more than informed opinion be formed in sociological theorizing?  Does this imply that one can ultimately achieve only uncertainty?

As noted above, "uncertainty" may substantively mean something different in various dimensions.  Therefore, we need a definition which leaves room for variation in the substantive meaning of uncertainty, i.e., a definition which is analytically independent of any system of reference. A definition without reference to a system, however, has to be content-free, i.e. a mathematical definition.[xvi]

In 1948, Shannon provided us with such a definition of "uncertainty" as part of the mathematical theory of communication [51]. Shannon defined "information" as the uncertainty contained in a finite sequence of signals or, more generally, in a distribution. Whether one should call this quantity "information" has been heavily debated (e.g., [3, 6, 56]). But more important than these semantic problems, was Shannon's equation of the concept with probabilistic entropy [18]. In contrast to thermodynamic entropy, however, the probabilistic uncertainty is defined yet content-free, i.e., it is still open to substantive specification.

Thermodynamic entropy is a measure of disorder among molecules in thermodynamics, and it can also be used to describe the direction of time in evolutionary processes (e.g., [7, 8, 56]). In the social sciences, however, one is usually not interested in the non-equilibrium thermodynamics of a physico-chemical system, but in the development of uncertainty, disorder, and complexity in social systems. Thus, the uncertainty refers to a different substance, and it can be reflected only by a different theory of communication.

## 2.2. **The probabilistic interpretation of communication**

How can substances communicate if there is no pre-established harmony and synchronicity? The envisaged generalization of concepts like "entropy" and "communication" to the dynamics of systems other than the physico-chemical one requires a further reflection on the assumptions contained in the mathematization of physics. As noted, the concept of communication is much older than the thermodynamic concept of entropy [4] or its probabilistic interpretation in the mathematical theory of communication [51]. Descartes and Huygens, for example, had to assume that "motion" (momentum and energy) is communicated in a collision in order to be conserved, and thus they discussed this conservation in terms of the "laws of communication of motion."[xvii] I showed above that Huygens gave the Cartesian concepts a physical interpretation. I shall now use the example of the collision in a classical system to infer the probabilistic concept of communication from this older notion of communication.

In a system of colliding balls momentum and energy have to be conserved, and thus to be communicated upon collision. As we know nowadays, the efficiency of the communication of momenta in a physical realization depends on the amount of (free) energy which dissipates as thermodynamic entropy. The ideal communication of momenta and kinetic energies of the colliding balls is thus dampened by this dissipation. When the physical realization approximates the ideal case, the thermodynamic entropy vanishes, but the redistribution of momenta and energies at the macro-level becomes more pronounced (since there is less dissipation). Correspondingly, the message that the collision has taken place contains a larger amount of Shannon-type uncertainty. Thus, the two types of entropy can vary independently: the one may increase and the other vanish in the same event. The reason for this independence is that the systems of reference for the two entropies are different: thermodynamic entropy refers exclusively to the distribution of, for example, momenta and positions among molecules, while the reference system for probabilistic entropy in this case is the system which conserves macroscopic momenta and energy. *Thermodynamic entropy is generated only in the special case where the communication has the physico-chemical system as its substantive reference.*[xviii]

Shannon's probabilistic definition of entropy enables us to develop a content-free definition of communication systems which operate by processing distributions. In the example above, the macroscopic energy system communicated in terms of the kinetic energies of (billiard-type) balls, the momentum system in terms of momenta. Social systems communicate in terms of means of social communication (e.g., discourse, money, etc.); human bodies communicate in terms of hormones and neural potentials. In these cases the probabilistic entropy is defined with reference to systems other than the physico-chemical one.

In summary, the translation of contingent uncertainty into mathematical clarity by Descartes has been generalized by Shannon to the understanding of a contingency as a probability distribution. Like the uncertainty in the act of doubt, the mathematical awareness of a probabilistic event cannot be given a substantive meaning internally by this theoretical system; it needs an external reference. However, the external reference again need not be *physical* existence. In systems other than the physical one, other quantities than "motion" may have to be conserved, and therefore communicated.

For example, in classical chemistry a mass balance for each element involved in the reaction is assumed. In this case, the atoms of the elements are redistributed. One can express the communication of any redistributed quantities as a message which contains information, and thus in terms of probabilistic entropy. The systems (and subsystems)[xix] are different with respect to the quality of *what* is being communicated, not with respect to the generation of probabilistic entropy. If the system under study generates probabilistic entropy with respect to two communications (e.g., on the occasion of a collision with respect to energy and momentum), a probabilistic entropy is generated in each dimension of relevance. In general, *the number of dimensions of the information in the message that the* *event happened is equal to the number of systems of reference for the information.* Each system of reference adds another quality to the uncertainty, and therefore another dimension to the communication.

Thus we arrive at a general formulation of the problem noted by Huygens that the dimensionality of the uncertainty has to be specified.  When Huygens refered to mathematical space *and* physical extension, he hypothesized two dimensions (i.e., mathematical *a priori* knowledge and physical uncertainty), where Descartes had hypothesized only one dimension, in which clarity consequently can substitute for uncertainty.  If, for example, in a chemical reaction three (qualitatively different) elements have to be balanced in terms of their respective total mass, the message of this event will analogously contain a three-dimensional uncertainty.

Information is never free-floating, but necessarily itself processed within a contingent communication system.  The communication systems are delineated in terms of what they communicate.  Whatever they communicate is redistributed in the communication, and this redistribution is in itself a message which is sent to all the communication systems with which this system can communicate externally.  In a single communication, i.e., by its contingent operation, the system communicates internally that it has reached a new state, and externally to all coupled systems that this contingency has happened in their environment. Analogously, the receiving systems can only receive the message by operating, and thus by redistributing their own information contents.  Cycles of communication are thus generated. The complexity increases rapidly (i.e., with the exponent of the number of systems)$^{xx}$ unless the systems are also able to (self-)organize the information.

What are the conditions under which communication systems can also organize the uncertainty, either among one another or internally?  In other words: what are the conditions under which networks can retain and organize information?  As noted, some systems are conservative, i.e., the number of elements which can be communicated is fixed.  In general, the number of elements ($n$) which a system contains sets a limit to the information which the system can hold.   One may also express this as the maximal entropy (*viz.*, equal to log(*n*)).  As noted above, the number of elements in systems can be multiplied by adding other systems of reference to the communication, and thus by increasing the number of dimensions in the information (*n* x *m*).   Furthermore, open systems like social communication systems can be defined only in terms of the communication, and consequently these systems have uncertain boundaries.   Each additional node of the network *n* adds (*n - 1*) possible links.   In general, when the number of elements increases more rapidly than the information content of the system, the redundancy which can be defined as the complement of the information content also increases.   Thus, the addition of new dimensions or new elements can lead to a *relative* decrease of the probabilistic entropy contained within the system.[xxi]   In other words, the uncertainty can be reduced within the system either by increasing the internal complexity or by growth.

The maintenance of the system is a balanced outcome of its necessary production of (probabilistic) entropy by operating, and this capacity to organize the uncertainty within the system.[16, 55]   Self-organization [46] or *autopoiesis* [41] can only be achieved by communication systems which are able to reflexively vary the *organization* of the uncertainty along the time dimension.   In other words, self-organizing systems reconstruct their histories so that they can face their future in terms of expectations.   Note that this reflexive capacity can never be observed directly, but only hypothesized as an internal mechanism of the system(s) under study.[38]

In general, communication systems develop through processing, i.e., by redistributing whatever they communicate.   With respect to this processing one can distinguish between self-referentiality (the internal processing of the message that the *a priori* distribution of the substance of communication was changed into the *a posteriori* one), and external referentiality to all systems of reference.   On the one side, the number of reference-systems determines the dimensionality of the information content of the self-referential update.   On the other side, the frequency of the update sets the system's clock.   Note that this frequency can be multi-variate, and thus be a frequency distribution, i.e., a spectrum. The clocks tick with a variety of speeds. There is no *a priori* reason for harmony: communications are in principle asynchronous.

Thus, in addition to providing a potentially multi-variate environment for one another, the communication systems constitute each others' environments in terms of time. To the extent that communication among systems is sustained, the systems also have to communicate frequency distributions in the time dimension. However, time is not a normal variate. This further complicates the analysis.

2.3. **An example of a multi-system communication**

Before extending the analysis in the time dimension, let me illustrate this abstract conceptualization by elaborating on the simple example of a telephone conversation as a communication with relevance for two qualitatively different systems, i.e., the social system and the telephone network.

First, the contingency of a telephone conversation can be analyzed in terms of physical currents through a network which can be mathematically modelled. The social communication in a telephone call, however, remains external to the mathematics of the propagation of signals through the lines. Nevertheless, the social communication system and the telephone system interact in this single event. By operating both systems change as a consequence of the interaction. (Of course, the sending and the receiving systems are also involved.)

The social system and the telephone network, however, were not *a priori* in harmony. No perfect deity is involved, but only a couple of engineers who have done their utmost to make the telephone system function. As Latour ([31], at p. 188) noted: "There is no preestablished harmony, Leibniz notwithstanding, harmony is *post*established locally through tinkering." However, a user may fail to establish the connection: each communication system remains failure-prone in the interaction. Additionally, each of the two systems, while related to the other system in the unique event of this historic phone call, does not contain nor acquire full information about the contingent boundaries of the other system through these interactions.  In general, the two systems remain *virtual* for one each other while interacting.  They can observe one another only through the "lens" of the interaction.

Although virtual, the two systems are *not transparent* for one another: it makes a difference whether people communicate by telephone or through other means of communication, and it may make a difference for the telephone line whether it transported data or voice-input (e.g., in terms of costs of the transmission).  In the interaction, the two systems "disturb" one each other, but they do not delimit each other.  Thus, they are each other's environment only in the specific sense of having a communication window on each other.  Note the difference here from the concept of the relation between system and environment in, for example, biology.[xxii]

In summary, the two systems disturb each other in the event of the historical interaction.  The disturbance is a contingent event, since it could have been otherwise.  It is a single contingency, but it has a different relevance for each system of reference.  Within each system the uncertainty in the event can be evaluated with reference to the self-referential contingency within the respective system.  The contingency of the one system is underdetermined by the other, since it is not delineated from it as such, but only in the interaction.  Analogously, the time-horizon in the other system is also not delimited by the interaction.  The systems communicate in relation to one another autonomously like Leibniz' monads, but they are contingent!  However, since they cannot fully perceive each other's contingency, the systems are autonomous centres of control in relation to one another, and only on this basis can they interact.  In this interaction, it is not clear for each system which systems interact, since each system only contains its own contingency, although each system is partially also informed about the interacting systems by the interaction.

However, only systems which can reflexively reconstruct, in addition to being part of a (relational) construction, can produce expectations.  In the reconstruction, each system has no other source of information about the *possible* interactions in the communication with other systems than the information which it can retrieve from its own history. But the system can only generate knowledge internally from this uncertainty, if it is capable of storing information about its previous states, and if it is additionally capable of taking this information reflexively into memory. If so, it may position itself historically, and in relation to the multi-dimensional space of systems of reference at each moment in time, and thus produce meaning in a second-order cybernetics. Reflexive reconstruction requires the capacity of the system to take the contingent self-referentiality of the system's history into memory. Obviously, human *cogito*'s are (among)[xxiii] systems which can act reflexively.

As noted, Huygens reconstructed his experience within his contingent *cogito* differently from Descartes. However, if a *cogito* expects that another system is a relevant (i.e., disturbing) environment, how many negative instances does the *cogito* need in order to revise this hypothesis? In other words: how frequently does it internally update this reconstruction in relation to the ongoing construction at the operational level? Additionally, one may raise the question of whether social systems or theoretical knowledge systems are not only constructed, but are also reconstructive, and whether they are also able to update in a second-order cybernetics. However, this raises further questions about the dynamics of distributed memory management, since the memory function of social systems is operationally located in human beings [36, 39].

2.4. **Extension to the time dimension**

Remember that we arrived in the first part of this study at the conclusion that without further demarcation, the reflexive communication system contains only information in the time-dimension about the frequency of its self-referential update, and it knows itself to be contingent. However, it can determine what it communicates substantively only with reference to an environment; and it can only receive information from the environment insofar as the environment consists of other communication systems. Thus, this notion of a system is yet content-free: the contingency refers only to its finite character, its being sequenced in time, and its being the yet unspecified substance of a communication system among other communication systems.

The special character of time as a variate of a communication system was only recently made a focus of methodological reflection in the social sciences. If two (or more) systems communicate parts of their expected information content by co-varying, they will usually have changed *ex post* when compared with the situation *ex ante*. The co-variation represents the interaction, while the remaining variances on both sides represent the respective continuities. In other words, one expects both continuity and change in the systems under study. The remaining parts of the variances co-vary with a previous state of the system (i.e., over time), and are therefore "auto-correlated." If variances are auto-correlated, then so are their error terms, and this violates a central assumption in regression analysis [5]. Furthermore, a multi-variate system is expected to develop differently from a set of non-coupled elements. Since each two or more of these elements may form a system (or a subsystem within a system), the number of possible expectations for future behaviour increases exponentially with the number of elements, and thus the inductive analysis rapidly becomes over-complex.[xxiv] The methodological statement that time-series data should not be used for regression analysis without correction for auto-correlation in the data, means in qualitative terms that change in the multi-variate data can only be assessed on the basis of an hypothesis for the delineation of the self-referential system(s) that exhibit the observable interaction(s).

Qualitative sociologists, therefore, are right when they state that existing statistical models in the social sciences cannot cope with the complexities of social developments in the historical dimension.  Social science statistics is most sophisticated in addressing problems of multi-variate analysis, but in a dynamic design there are shortcomings with respect to the combination of the multi-variate and the time series perspective.[34]  How can an historical series of events be assessed for its significance in relation to the range of developments which might have occurred?

The common solution on the qualitative side is to take the historical axis as a sort of independent variable, to which all other developments are then discursively "regressed" in a narrative.  This solution, however, is irreflexive with respect to the time dimension; one should not assume that there exists one single (i.e., historical) time.  Time can only be defined with reference to a clock, and a clock can only be a system's clock.  System clocks, however, may tick according to a spectrum of different frequencies.

In general, clocks of contingent systems are expected to be asynchronous.  There is no *a priori* reason why the various periodicities should be the same for different systems, i.e., why different systems should operate synchronously.  Synchronization is a local event, which requires explanation.  For example, it is only a consequence of the rotation of the earth that many systems on earth happen to be updated daily.  In addition to whatever information may be communicated, systems with a history must also update mutually, and occasionally must synchronize in the time dimension.

Communication systems generate variation for each other by redistributing their configurations.  A reflexive analyst may be able to use the observable interactions as information about the systems under study, and about their development.  The systems are not observed, but remain expectations.  Thus, in order to solve the problem of "auto-correlation" in the data, one has to reverse the reasoning: auto-correlation is not first to be corrected for on the basis of an assumed ideal case, but systems can only develop over time self-referentially, i.e., with reference to themselves at a previous moment. If the (reconstructive) analysis leads to the conclusion that the variations are not self-referential--i.e., not auto-correlated--this may indicate a special case where the systems under study changed so importantly that a completely different system emerged (cf. [35]). Alternatively, the interacting systems may not have been correctly hypothesized.

In general, communication in the time dimension is an event like all other communications. What is communicated is a frequency distribution (i.e., a spectrum).[54] Analogously to communication in other dimensions, some communication systems are only able to communicate this information, others are able to store it, and specific ones are able to reflect upon it and give it an interpretation. Note that communications are discrete events, and that thus continuous time is an idealization by the reconstructive system. Consequently, one should be cautious in using differential calculus for the reconstruction because of the assumptions contained in it concerning the limit transition to continuous time.[xxv] If the post-modern understanding were to assume a standard clock, it might be caught eventually within the very cosmology which it wished to overcome. Synchronization among systems always requires explanation.

3. **Towards a general theory of communication**

In analogy to the probabilistic interpretation of entropy, and the consequential definition of time in terms of spectra of frequencies, one can give a probabilistic interpretation to concepts in physics which build on the notion of entropy. However, since codified knowledge in physics is logically consistent, other concepts of modern physics can also be given a probabilistic, i.e., non-physical, interpretation in a mathematical theory of communication.

How should one understand a probabilistic interpretation of concepts and laws from physics? An insightful access is provided by using those concepts which, like the Boltzmann equations, rely heavily on the concept of entropy.    From the probabilistic interpretations of these laws and concepts one can derive content-free (mathematical) theory, which can subsequently be given meaning with reference to systems other than the chemico-physical one.

In practice, computer scientists and cognitive scientists have already begun to investigate the usefulness of Boltzmann equations for modelling complex network problems (e.g., [54]).    For example, if a system tends to be in discrete states, the probability of finding the system in each of these states is not different in the computation than the probability of finding an electron in the various orbits which are allowed in an atom.    (These discrete states may also be considered as "attractors.")    Thus, we have the rich mathematical apparatus of physics at our disposal for studying systems which can be described in terms of probability distributions.

Let us take the concept of probabilistic temperature as an example.    At prevailing probabilistic temperatures one observes both the (self-)organization of systems (i.e., storage of probabilistic energy) and their generation of entropy in interactions (i.e., dissipation of probabilistic energy).    However, if one "freezes" the systems, one removes the factor of dissipation by bringing all systems to their lowest energy states (according to the Boltzmann equation).    In chemical physics, we know this state as, for example, crystalline.    The attractors can be sorted separately, since they "peak" against one another in the observation at extremely low probabilistic temperatures.    Note, however, that a probabilistic temperature is not a physical temperature, but a content-free concept which can only be given meaning with reference to a system (or a system of systems).

The range of applications of these probabilistic simulations is fascinating: on the one hand, in cognitive psychology attractors are constructed by training computer networks, e.g., for pattern recognition (so-called "Boltzmann-machines"; cf. [12, 20, 49]).    On the other hand, for example, Kuhn's [26] concept of "paradigms" provided us with a mental model of the possibility of attractors in the social system: the paradigm not only controls what is communicatable within it, but also shapes a social boundary between those who are "inside" and "outside" the relevant scientific community.  Analogously, regimes can be considered as the higher-order attractors of interactions among localizable trajectories and socially distributed learning processes.[xxvi]

The extension of concepts from physics to non-physical realms may sound at first like positivism, but this is not positivism.   First, we did not impose the model of physics normatively upon the other sciences, but we used the results of modern physics reflexively for the understanding of systems other than the chemico-physical one by first giving the concepts a different (i.e., probabilistic) interpretation.   Other systems are, among other things, much more complex than the chemico-physical one in terms of what is being communicated.   For example, in a simple biological system a large number of mass balances are already involved.   In psychological systems, people process feelings and thoughts, which are most difficult to operationalize so that they can be externally observed.   In social systems, people communicate by using language and symbolic media of communication. The nature of these communications, i.e., their operationalization, can only be specified by theorizing at the relevant systems level.   Thus, the observable interactions should not be taken as the units of analysis.   They are the phenotypical results which challenge the theoretical understanding for specifying the genotypical mechanisms.[28, 37]   A general theory of communication can be expeted to guide us with respect to the algorithmic modelling of the interactions among the so specified communication systems, and to provide us with the mathematics for explaining their probabilistic behaviour over time.

4. **Discussion and Conclusions**

The embeddedness of the knowing subject in what it wants to investigate pointed to the reconstructive and reflexive nature of human knowledge. However, in the epistemological reflection one originally focussed on the question of what specific contingency meant for the development of the whole, which was itself specified in terms of a transcendency. In the natural sciences, for example, one has assumed that one could abstract from the specific positions of people with reference to the natural environment by using the concept of a transcendental subject.

In relation to society, or more generally with reference to social systems, this metaphysically warranted assumption of commonality disintegrated in the 19th century (cf. Marx). The claim of an objective meta-position is nowadays untenable in the social sciences, since it is, for example, irreflexive to the bias which is necessarily brought into the analysis by initial assumptions. Whether this bias is a class position, a male bias or a wish to dominate the discourse (cf. Foucault) is secondary. The primary point is that a theoretical system reconstructs the social system from a particular point of view.

The mere formulation of the objective of general theory, therefore, may seem an invitation to obscurity for those social scientists and philosophers who deny the possibility of general theory on normative and sociological grounds. Indeed, the issue of general theory in sociology emphatically raises the issue of the position of the observer, and of the theorist's own historicity. Since Max Weber this complex of issues has been debated in terms of the (voluntaristic) theory of action [42]. However, does the historicity of an individual act destroy *a priori* the possibility of reconstructing society by using a theoretical model? In my opinion, the problem of historicity specifies only one criterion for a theoretical model, namely that it should be able to account for historicity. Additionally, theory should be able to cope with its own historical contingency reflexively, i.e., to understand itself in terms of a reconstruction.

Of course, the specification of a general theory of communication goes beyond the scope of this study.[37, 38]   The crucial point, however, is that neither the substance under study nor the scientific communication system should be considered as spatial extensions (e.g., domains) only; all communication systems contain contingency in *four* dimensions, i.e., in space and time.   Observable stability is the special case in which one has to assume "the continual reiteration or propagation of an already presupposed effort and counter-effort" (Leibniz)[xxvii] or--as we would now say--of a positive feedback.   Thus, an observation can only be informative with reference to an expectation, but the theoretical expectation is embedded in a system of expectations.   One may wish to close the system at either level, but the closure is temporary and can be deconstructed.

Newton and Leibniz understood that substance should be considered not as extension, but as force or action.   However, they stabilized their theoretical apparatus by basing it on *a priori* foundations.   On the one hand, these scholars were able to entertain concepts like "gravity" and "acceleration", since the calculus provided them with the concept of a second derivative.   Obviously, if one wishes to explain events in a hyper-space of space and time, one eventually needs to supplement the geometrical measurement with an algebraic understanding.[cf. 27]

On the other hand, this conclusion has consequences for those sciences that have hitherto relied on geometrical narratives for their understanding [17, 54].   In a second-order theory the theoretical apparatus is itself reflexive on its contingency; it knows itself to be a communication system among other possible communication systems, subject to continual changes.   But since both the data and their interpretation are in flux, one additionally needs an algebraic model for the theoretical self-understanding.   This next-higher-order complexity in comparison to Newtonian physics calls for the interpretation of results in algorithmic "computerese" as a higher-order extension of the "natural" language that has used mainly geometrical metaphors.[2, 28, 37]

A general theory of communication adds to Shannon's mathematical theory of communication the concept of systems of reference, and the non-equilibrium perspective. With respect to the systems of reference, one needs special theories (by definition). The non-equilibrium perspective enables us to model evolutionary processes such as paradigm developments, lifecycles, etc. The scientific model, however, remains reconstructive, and therefore part of a cultural evolution. The reflexive awareness of this methodological status is the one important aspect in which communication theory differs from biological evolution theory. The latter hypothesized "natural selection" by the environment as an external principle which independently organizes a variety of taxonomic data. Evolution theory then allows us, for example, to define "missing links" in the evolutionary data, and it guides us in searching for unambiguous evidence of these instances. Reconstructions, however, provide us with alternative hypotheses concerning what has guided the system(s) under description. The alternative hypotheses may describe various aspects of learning, and the consequent emergence of patterns of behaviour and communication, which may then begin to act as selection mechanisms.

The higher-order selection environments do not have to develop synchronously with the systems under study. A second-order cybernetics between selection and stabilization can be assumed (e.g., [30, 38, 40]). *Evolution theory is then the special case in which the (natural) environment is considered the single determining factor for selection.* Sociological data, however, exhibit a multitude of dynamics, and the various systems are only hypothesized systems of reference ("attractors") instead of a single evolution. Thus, in relation to biology, the socio-cultural perspective adds reflexivity to the theoretical inference.[37, 40] While in other sciences it may have been fruitful to take either variation or selection as predetermined by "Nature" as a cosmologically warranted system of reference, sociological theorizing requires a reflexive awareness of the variance and historicity of both dimensions.

**Notes**

i. Huygens [22], Vol. XXI, at p. 541. See also: [14], at p. 37.

ii. Huygens speaks of his own method as consisting of *experientia ac ratione*, that is, proceeding with experience and reason.[14]

iii. [22], Bk. I, at pp. 9-10. See: [14], at p. 38.

iv. Whether this is still the case for quantumphysics is a separate issue. For this discussion, see for example [45].

v. Huygens [22], Vol. XIX, at p. 325. See also: [15], at p. 131.

vi. Letter of November 18, 1690. ([22], Vol. IX, at p. 538.)

vii. Note here the Cartesian notion of God: before delineation, i.e., in its self-referential intimacy, the contingency is exclusively defined in relation to its transcendency, i.e., in relation to God.   Since the definition is internal to the specific *cogito*, this implies a self-referential relation to a personal God, who is present in the reflection.   In this sense, the Cartesian *Ego* reflects the Protestant revolution.

viii. Praise in the opening choir of Bach's cantata *Actus Tragicus* (1707).

ix. [32]; translated from the German edition: [33], at p. 269.

x. "Christian (...) qu'estant en l'estat où il se trouve, dans lequel il devroit comme envisager de pres l'immortabilité, il s'amuse à la controverter comme une question problematique pour et contre." Letter of 22 May 1670 by brother Lodewijk Huygens to the father, Constantijn Sr. ([22], Vol VII, at p. 22.)

xi. The preface to the first edition gives May 8, 1686 as the date.

xii. Huygens had betrayed the Dutch Republic when French armies had attacked and almost destroyed it in 1672.    Notably, he had dedicated his *Horologium Oscillatorum* in 1673 with the following opening sentence:    "We are especially indebted to France, Oh Great King, for the rebirth and restoration of geometry in this century."    For his glorious role in Paris, see for example: [52].

xiii. "I have been amazed that Huyghens and Newton assume the existence of empty space.    However, this can be explained from the fact that they have persisted to discuss in geometrical terms.    More astonishing is it still for me that Newton has assumed an attraction which does not work by mechanical means.    When he states with respect to this issue that the bodies attract one another in terms of gravitation, then should this not be discarded--at least, with respect to the observable interactions among the large bodies in our world system--although it seems that Huyghens also does not completely agree with this." (Leibniz in a letter to Bernouilli, 1698; translated from the German edition [33], at p. 371.)

xiv. The Kurfürst of Prussia, Friedrich I, who was later to be crowned as king *Fredericus Rex*, was a nephew of king William of Orange.    His mother Louise Henriette was a daughter of Frederik Henderik, Prince of Orange, who had relied heavily on the services of Huygens' father Constantijn Sr.    The princess was two years older than Christiaan Huygens, and as children they were raised in the same circles in The Hague.    Note also that Friedrich's wife, the later Queen Sophie Charlotte, was herself a philosopher.    She was a patroness of Leibniz (who lived in Hannover), and founded the Akademie der Wissenschaften in Berlin upon his instigation in 1700.

xv. Leibniz (1695) noted that otherwise "the souls (would) remain without purpose in a chaos of inextricable matter" ([34], at p. 262).

xvi. Since mathematics can also be one of the systems of reference, one may also wish to call this a meta-mathematical definition (cf. [20]).

xvii. "Within the framework of the Cartesian program, laws of motion ought to be laws of communication of motion expressed in measurable quantities." ([14], at p. 73.)

xviii. The Szilard-Brillouin relation shows that in this case only a very small part of the thermodynamic entropy ($S$) is probabilistic entropy ($H$). See also: [12], at p. 60.

xix. At this level of generality, one is not able to distinguish among systems and subsystems.

xx. When complexity increases not with the power of $n$ (i.e., $n^k$), but with the exponent of $n$ (i.e., $\exp(\beta n)$), the problem can be non-polynomial complete, and therefore, becomes uncomputable in practice. See, for example: [12, 45].

xxi. The number of possible states of the network increases with the exponent of the number of its nodes.

xxii. However, the concept of self-organization, and its implications for the relations between systems and environments, is often discussed also in relation to (biological) evolution theory. See, among others: [25, 29].

xxiii. Since the systems and their operations were yet defined as content-free, the human being is formally a specification (cf. [30, 38]). Additionally, one has to specify what is reflected in the reflection (e.g., thought, feelings, etc.) and in terms of what it is reflected.

xxiv. Correspondingly, there are no auto-regressive (AREG and ARIMA) models for multi-variate data available, but only for uni-variate trendlines. If one wishes to predict the behaviour of a system of variables, one has to define a systems variable at the aggregate level, but then one risks losing perspective on how the variances within the system change. See also: [34].

xxv. Although the analyst may wish to use them for pragmatic reasons, the application of Shannon's

formulas to continuous distributions is theoretically more problematic than their application to discrete ones. See also: [55], at p. 74.

xxvi. In a study of the management of natural resources, Allen [1] found two attractors in the parallel simulation of the hyperbolic curve of fish against fishing boats. In formal terms, this curve is similar to a traditional production function with capital and labour along the axes (cf. [11, 37, 44, 50]).

xxvii. Quoted from Leibniz' *Specimen Dynamicum* by [48], at p. 251.    See also: [58].